\documentclass[aps,prl,twocolumn,groupedaddress,draft,showpacs]{revtex4}

\begin{document}

\title{Parity and Time Reversal in the Spin-Rotation Interaction}

\author{G. Papini}
\email[]{papini@uregina.ca}
\thanks{Research supported by the Natural Sciences and Engineering Research Council of Canada.}

\affiliation{Department of Physics, University of Regina, Regina, Sask. S4S 0A2,
Canada\\International Institute for Advanced Scientific Studies, 84019 Vietri sul Mare
(Sa), Italy.}

\date{\today}

\begin{abstract}
A recently reported discrepancy between experimental and theoretical values of the muon's
$g-2$ factor is interpreted as due to small violations of the conservation of $P$ and $T$
in the spin-rotation coupling. The experiments place an upper limit on these violations
and on the weight change of spinning gyroscopes.
\end{abstract}


\maketitle

The spin-rotation effect described by Mashhoon \cite{mashhoon} attributes an energy
$-\frac{\hbar}{2}\vec{\omega}\cdot \vec{\sigma} $ to a spin-$\frac{1}{2}$ particle in a
frame  rotating with angular velocity $\omega$ relative to an inertial frame. Identical
results have been derived directly from the Dirac equation by means of the tetrad
formalism \cite{hehlni,caipapini,singhpapini}. The effect extends our knowledge of
rotational inertia to the quantum level and  violates the principle of equivalence
\cite{mashhoon1} while preserving invariance under $P$ and $T$. It has physical and
astrophysical implications \cite{lloydcaipap,caipapini,papini,papini1} and also plays a
fundamental role in precise measurements of the $g-2$ factor of the muon \cite{papini1}.

Recently, a discrepancy $a_{\mu}(exp)-a_{\mu}(SM)=43\times 10^{-10}$ has been observed
\cite{brown} between the experimental and standard model values of the muon's anomalous
$g$ value, $a_{\mu}=\frac{g-2}{2}$. This discrepancy can be used to set an upper limit on
$P$ and $T$ invariance violations in spin-rotation coupling.

The possibility that discrete symmetries in gravitation be not conserved has been
considered by some authors \cite{schiff,leitner,dass,almeida}. Attention has in general
focused on the potential (in units $\hbar=c=1$)
\begin{equation}\label{1}
U(\vec{r})=\frac{GM}{r}\left[\alpha_{1}\vec{\sigma}\cdot
\hat{r}+\alpha_{2}\vec{\sigma}\cdot \vec{v}+\alpha_{3}\hat{r}\cdot( \vec{v}\times
\vec{\sigma})\right],
\end{equation}
which applies to a particle of generic spin $\vec{\sigma}$. The first term, introduced by
Leitner and Okubo \cite{leitner}, violates the conservation of $P$ and $T$. The same
authors determined the upper limit $\alpha_{1}\leq 10^{-11}$ from the hyperfine splitting
of the ground state of hydrogen. The upper limit $\alpha_{2}\leq 10^{-3}$ was determined
in Ref.\cite{almeida} from SN 1987A data. The corresponding potential violates the
conservation of $P$ and $C$. Conservation of $C$ and $T$ is violated by the last term,
while (\ref{1}), as a whole, conserves $CPT$. There is, as yet, no upper limit on
$\alpha_{3}$. These studies are extended here to the Mashhoon term.

The $g-2$ experiment involves muons in a storage ring \cite{farley}. As the muons decay,
the angular distribution of those electrons projected forward in the direction of motion
reflect the precession of the muon spin along the cyclotron orbits.

Assume that all quantities in the effective Hamiltonian are time-independent and referred
to a left-handed tern of axes comoving with the muons and rotating about the $x_{2}$-axis
in the clockwise direction of motion of the muons. The $x_{3}$-axis is tangent to the
orbits and in the direction of the muon momentum. The magnetic field is $B_{2}=-B$. Of
all the terms that appear in the Dirac Hamiltonian, only the Mashhoon term couples the
helicity states of the muon. The remaining terms contribute to the overall energy $E$ of
the states and the corresponding part of the Hamiltonian is indicated by $H_{0}$
\cite{papini1}.

Before decay, the muon states can be represented by
\begin{equation}\label{2}
\mid\psi(t)>=a(t)\mid\psi_{+}>+b(t)\mid\psi_{-}>,
\end{equation}
where $\mid\psi_{+}>$ and $\mid\psi_{-}>$ are the right and left helicity states of the
Hamiltonian $H_{0}$ and satisfy the equation
\begin{equation}\label{3}
H_{0}\mid\psi_{+,-}>=E\mid\psi_{+,-}>.
\end{equation}

Assume now that the coupling of rotation to $\mid\psi_{+}>$ differs in strength from that
to $\mid\psi_{-}>$. Then the Mashhoon term can be altered by means of a matrix
$A=\left(\matrix{\kappa_{1}&0\cr 0&\kappa_{2}\cr}\right)$ that reflects the different
coupling of rotation to the two helicity states. The total effective Hamiltonian is
$H_{eff}=H_{0}+H'$, where
\begin{equation}\label{4}
H'=-\frac{1}{2}A \omega_{2}\sigma_{2}+\mu B\sigma_{2},
\end{equation}
$\mu=(1+a_{\mu})\mu_{0}$ represents the total magnetic moment of the muon and $\mu_{0}$
is the Bohr magneton. A violation of $P$ and $T$ in (\ref{4}) would arise through
$\kappa_{2}-\kappa_{1}\neq 0$. The constants $\kappa_{1}$ and $\kappa_{2}$ are assumed to
differ from unity by small amounts $\epsilon_{1}$ and $\epsilon_{2}$.

The coefficients $a(t)$ and $b(t)$ in (\ref{1}) evolve in time according to
\begin{equation}\label{5}
i\frac{\partial}{\partial t}\left(\matrix{a(t)\cr b(t)\cr}\right)=M\left(\matrix{a(t)\cr
b(t)\cr}\right),
\end{equation}
where
\begin{equation}\label{6}
M=\left(\matrix{E-i\frac{\Gamma}{2}& i\left(\kappa_{1}\frac{\omega_{2}}{2}-\mu
B\right)\cr -i\left(\kappa_{2}\frac{\omega_{2}}{2}-\mu B\right)&
E-i\frac{\Gamma}{2}\cr}\right),
\end{equation}
and $\Gamma$ represents the width of the muon. The spin-rotation term is off-diagonal in
(\ref{6}) and does not therefore couple to matter universally. It violates Hermiticity
\cite{wald}. It also violates $T$, $P$ and $PT$, as stated, while nothing can be said
about $CPT$ conservation which requires $H_{eff}$ to be Hermitian
\cite{kennysachs,sachs}. Because of the non-Hermitian nature of (\ref{4}), one expects
$\Gamma$ itself to be non-Hermitian. The resulting corrections to the width of the muon
are, however, of second order in the $\epsilon$'s and are neglected.

$M$ has eigenvalues
\begin{eqnarray}
h_{1}&=&E-i\frac{\Gamma}{2}+R \nonumber \\ h_{2}&=&E-i\frac{\Gamma}{2}-R,
\end{eqnarray}
where
\begin{equation}
R=\sqrt{\left(\kappa_{1}\frac{\omega_{2}}{2}-\mu
B\right)\left(\kappa_{2}\frac{\omega_{2}}{2}-\mu B\right)},
\end{equation}
and eigenstates
\begin{eqnarray}
|\psi_{1}>&=&b_{1}\left[\eta_{1}|\psi_{+}>+|\psi_{-}>\right],\nonumber \\
|\psi_{2}>&=&b_{2}\left[\eta_{2}|\psi_{+}>+|\psi_{-}>\right].
\end{eqnarray}
One also finds
\begin{eqnarray}
|b_{1}|^{2}&=&\frac{1}{1+|\eta_{1}|^{2}}\nonumber \\
|b_{2}|^{2}&=&\frac{1}{1+|\eta_{2}|^{2}}
\end{eqnarray}
and
\begin{equation}
\eta_{1}=-\eta_{2}=\frac{i}{R}\left(\kappa_{1}\frac{\omega_{2}}{2}-\mu B\right).
\end{equation}
Then the muon states (\ref{2}) are
\begin{eqnarray}
|\psi(t)>&=&\frac{1}{2}e^{-iEt-\frac{\Gamma t}{2}}[-2i\eta_{1} \sin Rt |\psi_{+}>+
\nonumber \\
         & &2\cos Rt
         |\psi_{-}>],
\end{eqnarray}
where the condition $|\psi(0)>=|\psi_{-}>$ has been applied. The spin-flip probability is
therefore
\begin{eqnarray}\label{9}
P_{\psi_{-}\rightarrow \psi_{+}}&=&|<\psi_{+}|\psi(t)>|^{2}\nonumber
\\&=&\frac{e^{-\Gamma t}}{2}\frac{\kappa_{1}\omega_{2}-2\mu B}{\kappa_{2}\omega_{2}-2\mu
B}\left[1-\cos 2Rt\right].
\end{eqnarray}
When $ \kappa_{1}=\kappa_{2}=1$, Eq.(\ref{9}) yields  \cite{papini1}
\begin{equation}\label{10}
P_{\psi_{-}\rightarrow \psi_{+}}=\frac{e^{-\Gamma t}}{2}\left[1-\cos\left(a_{\mu}
\frac{eB}{m}t\right)\right],
\end{equation}
that provides the appropriate description of the spin-rotation contribution to the
spin-flip transition probability. Notice that the case $\kappa_{1}=\kappa_{2}=0$ (no
spin-rotation coupling) yields
\begin{equation}\label{A}
P_{\psi_{-}\rightarrow \psi_{+}}=\frac{e^{-\Gamma
t}}{2}\left[1-\cos(1+a_{\mu})\frac{eB}{m}\right]
\end{equation}
and does not therefore agree with the results of the $g-2$ experiments. Hence the
necessity of accounting for spin-rotation coupling whose contribution cancels the factor
$\frac{eB}{m}$ in (\ref{A})\cite{papini1}.

 Substituting $\kappa_{1}=1+\epsilon_{1},
\kappa_{2}=1+\epsilon_{2}$ into (\ref{9}), one finds
\begin{equation}\label{11}
P_{\psi_{-}\rightarrow\psi_{+}}\simeq\frac{e^{-\Gamma t}}{2}[1-\cos
\frac{eB}{m}(a_{\mu}-\epsilon)t] ,
\end{equation}
where $\epsilon=\frac{1}{2}(\epsilon_{1}+\epsilon_{2})$. One may attribute the
discrepancy between $a_{\mu}(exp)$ and $a_{\mu}(SM)$ to a violation of the conservation
of the discrete symmetries by the spin-rotation coupling term in (\ref{4}). The upper
limit on the violation of $P,T$ and $PT$ is derived from (\ref{11}) assuming that the
deviation from the current value of $a_{\mu}(SM)$ is wholly due to $\epsilon$. The upper
limit is therefore $43\times 10^{-10}$.

Some more information can be extracted from current $a_{\mu}$ data. One may in fact
assume that the coupling of rotation to the two helicity states of the fermion is
opposite. In this case the parameters have values $\kappa_{1}=1, \kappa_{2}=-1$. This is
the anti-Hermitian limit of the interaction. The oscillation frequency is then
\begin{eqnarray}\label{12}
R=\frac{1}{2}\sqrt{(2\mu
B)^2-\omega_{2}^2}&=&\frac{eB}{2m}\sqrt{2a_{\mu}+a_{\mu}^2}\nonumber \\ &\simeq&
\frac{eB}{m}\sqrt{a_{\mu}/2}
\end{eqnarray}
and Eq.(\ref{9}) gives
\begin{equation}\label{13}
P_{\psi_{-}\rightarrow \psi_{+}}\simeq \frac{e^{-\Gamma
t}}{2}\frac{a_{\mu}}{2+a_{\mu}}\left[1-\cos\left(\frac{eB}{m}\sqrt{2a_{\mu}}t\right)\right].
\end{equation}
Equations (\ref{13}) and (\ref{10}) differ in amplitude and frequency. In fact the
amplitude of (\ref{13}) is much smaller than that of (\ref{10}) while its frequency is
higher than that actually observed. The choice $ \kappa_{1}=1, \kappa_{2}=-1$ is not
therefore supported experimentally.

 It also follows from (\ref{4}), (\ref{5}) and (\ref{6}) that the weight of a rotating object depends on its
direction of rotation. The problem has been studied experimentally in \cite{hayasaka}. No
theoretical motivation for the study has ever been presented. An upper limit on this
effect can be obtained in the present model from (\ref{4}). The eigenstate energy
difference due to spin-rotation coupling is in fact
\begin{eqnarray}
-i\frac{\omega_{2}}{2}(<\psi_{-}|\sigma^{2}|\psi_{+}>&+&<\psi_{+}|\sigma^{2}|\psi_{-}>)=
\nonumber
\\ & & \frac{\omega_{2}}{2}(1+\epsilon).
\end{eqnarray}
The additional energy difference is therefore $ \frac{\epsilon \omega_{2}}{2}$, where $
\epsilon =-43\times 10^{-10}$. Returning to normal units, the corresponding decrease in
mass for a muon of positive helicity is $\Delta m=-\frac{\epsilon eB\hbar}{m c^2}\simeq
-3.1 \times 10^{-48} g$ and the decrease in weight is $g\Delta m\simeq 4\times 10^{-45}
dyne$.

The fraction of total rotational energy associated with the effect is $\epsilon /2$. If
one applies this result to all the particles of the gyroscope used in the experiment of
\cite{hayasaka}, then one finds that the energy difference of the two rotation states of
the body is at most $ \frac{\epsilon}{2}\frac{1}{2}I\omega^{2}\simeq 2.4$ $erg$,
corresponding to a change in mass $\leq 2.6 \times 10^{-21}g$ and a change in weight
$\leq 2.6 \times 10^{-18} dyne $, in agreement with the null experimental results of
\cite{faller,nitschke}.


\begin{thebibliography}{99}
\bibitem{mashhoon} B. Mashhoon, Phys. Rev. Lett. {\bf61}, 2639 (1988); Phys. Lett. A
{\bf143}, 176 (1990); A {\bf145}, 147 (1990); Phys. Rev. Lett. {\bf68}, 3812 (1992).
\bibitem{hehlni} F.W. Hehl and W.-T. Ni, Phys. Rev. D {\bf42}, 2045 (1990).
\bibitem{caipapini} Y.Q. Cai and G. Papini, Phys. Rev. Lett. {\bf66}, 1259 (1991);
{\bf68}, 3811 (1992).
\bibitem{singhpapini} D. Singh and G. Papini, Nuovo Cimento B {\bf115}, 223 (2000).
\bibitem{mashhoon1} B. Mashhoon, Classical Quantum Gravity {\bf17}, 2399 (2000).
\bibitem{lloydcaipap} Y.Q. Cai, D.G. Lloyd and G. Papini, Phys. Lett. A {\bf178}, 225
(1993).
\bibitem{papini} G. Papini, Proc. of the $5^{th}$ Canadian Conference on General Relativity and
Relativistic Astrophysics, Eds. R.B. Mann and R.G. McLenaghan, World Scientific,
Singapore, 1994, p. 10.
\bibitem{papini1} G. Papini, in "Advances in the interplay between quantum and gravity physics",
Eds. V. de Sabbata and A. Zheltukhin, Kluwer Academic Publishers, Dordrecht (in press);
arXiv:gr-qc/0110056; G. Papini, G. Lambiase, Phys. Lett. A {\bf294}, 175 (2002).
\bibitem{brown} H.N. Brown $et$ $al.$, Muon (g-2) Collaboration, Phys. Rev. Lett. {\bf86}, 2227
(2001).
\bibitem{schiff} L.I. Schiff, Phys. Rev. Lett. {\bf1}, 254 (1958).
\bibitem{leitner} J. Leitner and S. Okubo, Phys. Rev. {\bf136}, B1542 (1964).
\bibitem{dass} N.D. Hari Dass, Phys. Rev. Lett. {\bf36}, 393 (1976); Ann. Phys.(NY)
{\bf107}, 337 (1977).
\bibitem{almeida} L.D. Almeida, G.E.A. Matsas and A.A. Natale, Phys. Rev. D {\bf39}, 677
(1989).
\bibitem{farley} For example, see the review by F.J.M. Farley and E. Picasso, Advanced Series
in High-Energy Physics, vol. 7, Quantum Electrodynamics, Ed. T. Kinoshita, World
Scientific, Singapore, 1990, p. 479.
\bibitem{wald} Effective non-Hermitian interaction terms are at times considered in
general relativity. This in general reflects the fact that in the presence of a
gravitational field the number of particles in a certain channel can change. For example,
see R.M. Wald, Phys. Rev. D {\bf21}, 2742 (1980).
\bibitem{kennysachs} Brian G. Kenny and Robert G. Sachs, Phys. Rev. D {\bf8}, 1605
(1973).
\bibitem{sachs} Robert G. Sachs, Phys. Rev. D {\bf33}, 3283 (1986).
\bibitem{hayasaka} Hideo Hayasaka and Sakae Takeuchi, Phys. Rev. Lett. {\bf63}, 2701
(1989).
\bibitem{faller} J.E. Faller, W.J. Hollander, P.G. Nelson and M.P. McHugh, Phys. Rev.
Lett. {\bf64}, 825 (1990).
\bibitem{nitschke} J.M. Nitschke and P.A. Wilmarth, Phys. Rev. Lett. {\bf64}, 2115
(1990).
\end{thebibliography}
\end{document}